\documentclass[12pt,tightenlines,eqsecnum,floats,aps,amsmath,amssymb,nofootinbib,prd,showpacs]{revtex4}

\usepackage{amsmath,amssymb,amsfonts}
\usepackage{graphicx}
\usepackage{enumerate} 
\usepackage{colordvi} 

\def\be{\begin{equation}}
\def\ee{\end{equation}}
\def\ba{\begin{eqnarray}}
\def\ea{\end{eqnarray}}

\def\R{\mathbb{R}}

\def\k{\kappa}
\def\M{\mathcal{M}}
\def\H{\mathcal{H}}
\def\L{\mathcal{L}}
\def\ez{\,{}^o\! e}
\def\dd{\textrm{d}}

\def\f{\frac}
\def\p{\partial}
\def\h{\hat}

\def\t{\tilde}

\newcommand{\D}[0]{\rlap{$\mathrm{d}$}\hspace{0.3ex}\mathrm{d}}

\usepackage{enumerate}

\usepackage{colordvi}
\newcounter{mnotecount}[section]

\newcommand{\comment}[1]{}

\begin{document}
\preprint{\vbox{\baselineskip=12pt \rightline{IGC-08/04-}
}}
\title{Action and Hamiltonians in higher dimensional\\ general
relativity: First order framework}

\author{Abhay Ashtekar${}^1$}\email{ashtekar@gravity.psu.edu}
\author{David Sloan${}^1$}\email{sloan@gravity.psu.edu}
\affiliation{${}^{1}$Institute for Gravitation and the Cosmos, Penn State,
University Park, PA 16802, U.S.A.}

\begin{abstract}
We consider $d>4$-dimensional space-times which are asymptotically
flat at spatial infinity and show that, in the first order
framework, the action principle is well-defined \emph{without the
need of infinite counter terms.} It naturally leads to a covariant
phase space in which the Hamiltonians generating asymptotic
symmetries provide the total energy-momentum and angular momentum
of the isolated system. This work runs parallel to our previous
analysis in four dimensions \cite{aes}. The higher dimensional
analysis is in fact simpler because of absence of logarithmic and
super translation ambiguities.
\end{abstract}


\maketitle

\section{Introduction}
\label{s1}

The motivation and underlying ideas of this paper are the same as
that of our earlier analysis in 4 dimensions \cite{aes}. However,
for completeness, we will summarize the main points.

In most field theories the action depends only on fundamental
fields and their first derivatives. By contrast, the
Einstein-Hilbert action of general relativity depends also on the
second derivatives of the fundamental field, the space-time metric
$g$. As a consequence, stationary points of this action do not
yield Einstein's equations unless both the metric and its first
derivatives are kept fixed at the boundary; strictly we do not
have a well-defined variational principle. To remedy this
situation, Gibbons and Hawking \cite{gh,swh} proposed that we add
a surface term to the Einstein Hilbert action. We are then led to
\be \label{gh} S_{\rm EH+GH} (g)\, = \,  \f{1}{2\k} \left( \int_{\M}
R\, d^dV \,+\, 2\int_{\p\M} K \, d^{d-1}V + C \right)\, .\ee
Here $\k = 8\pi G$,\, $\M$ is a $d$-manifold representing an
appropriate portion of space-time, $\p\M$ its boundary, $R$ the
Ricci scalar of the metric $g$,\, $K$ the trace of the extrinsic
curvature of $\p\M$, and $C$ is an arbitrary function of the
metric $h$ induced on $\p\M$ by $g$.

Let us restrict ourselves to cases where $g$ has signature \,
(-,+,...,+), \, is smooth and globally hyperbolic. We will let $\M$
be the space-time region bounded between two Cauchy surfaces. If
$\M$ is spatially compact, by setting $C=0$, we obtain a
well-defined variational principle. However, in the asymptotically
flat case, it is well-known that this strategy has some important
limitations (see e.g., \cite{mm}). In particular, the action is
typically infinite even `on-shell', and indeed even when $g$ is the
Minkowski metric. To remedy this problem, Gibbons and Hawking
\cite{gh,swh} proposed an infinite subtraction: Carry out an
isometric embedding of $(\p\M, h)$ in Minkowski space, calculate the
trace $K_o$ of the extrinsic curvature of $\p\M$ defined by the
Minkowski metric, and set $C = -K_o$. However, the required
embedding does not always exist. For, in a $d$ dimensional
space-time the metric $h$ on the boundary $\p\M$ has $[d(d-1)/2] -
[d-1]$ degrees of freedom (after removing the diffeomorphism gauge)
while the choice of embedding provides a freedom worth only one
function on $\p\M$. Thus, even at this heuristic level, if $d \ge 4$
the freedom is not sufficient whence this infinite subtraction
procedure will not work for generic metrics $g$.

Over the last few years, a new set of proposals for infinite counter
terms $C$ have appeared in the literature. In particular, Kraus,
Larsen, Siebelink \cite{kls} have constructed a counter-term using a
(non-polynomial) function of the Ricci curvature of the boundary.
Mann and Marolf \cite{mm} have introduced a counter-term which is
closer to the spirit of the Gibbons-Hawking proposal. They replace
$K_o$ with the trace of a tensor field $\h{K}_{ab}$ which
generalizes the extrinsic curvature $K_{ab}^o$ of $\p\M$ with
respect to the Minkowski metric, used by Gibbons and Hawking, to
situations in which the boundary can not be isometrically embedded
in Minkowski space. Not only do these improved actions $S_{\rm imp}$
lead to well-defined action principles, but they also overcome
another limitation of the original proposal: Now $\delta S_{\rm imp}
=0$ at asymptotically flat solutions for \emph{all} permissible
variations $\delta$.

Since we are dealing just with \emph{classical} field theories with
\emph{smooth} fields, one might wonder if there is a way to avoid
infinite subtractions altogether and construct an action principle
which is manifestly finite from the beginning. The first goal of
this paper is to show in some detail that this is indeed possible if
one uses a first order framework based on orthonormal frames and
Lorentz connections. The second goal of the paper is to use this
action to construct a covariant Hamiltonian framework by keeping
careful track of boundary conditions.

If $d\ge 5$, the super translations and logarithmic translations
encountered in the $d=4$ \cite{aes} case are absent: the
asymptotic symmetry group is the Poincar\'e group even with the
`obvious' choice of boundary conditions. Heuristically, the
difference can be understood as follows. Since the solutions to
the Poisson equation in $d$-1 spatial dimensions fall off at
spatial infinity as $1/r^{d-3}$, to obtain non-zero mass at
spatial infinity, we have to allow metrics which approach a flat
metric only as $1/r^{d-3}$. If $d$=4, the resulting $1/r$ fall-off
makes the asymptotic symmetry group infinite dimensional because
of possible super translations \cite{ah,aa-ein,bs,ar} and
logarithmic translations \cite{aa-log}. The boundary conditions
needed to remove these potential asymptotic symmetries are subtle
(see \cite{aes} for a concise summary). In $d\ge 5$, the
$1/r^{d-3}$ deviation from flat space do not allow these extra
symmetries and the group of asymptotic symmetries is naturally the
Poincar\'e group. The only subtlety is that for the Hamiltonians
generating boosts to be well-defined, one has to impose a
`reflection symmetry' condition on the leading, $1/r^{d-3}$ part
of permissible metrics (more precisely, of ortho-normal frames).
Thus, overall, the higher dimensional analysis is considerably
simpler than that in four space-time dimensions.

The paper is organized as follows. In section \ref{s2} we
introduce the Lagrangian and Hamiltonian framework using the first
order framework. In section \ref{s3} we calculate the Hamiltonians
generating these Poincar\'e symmetries. Our approach has several
similarities to that used by Mann, Marolf, McNees and Virmani
\cite{mm,mmv,mmmv}. However, there are also two significant
differences. First, we use a first order ---rather than a second
order--- framework, thereby avoiding infinite counter-terms
altogether. Second, our boundary conditions are weaker in the
sense that we do not require that the space-time metric have an
asymptotic `Beig-Schmidt expansion'. On the other hand we require
that the coefficient of the leading, $1/r^{d-3}$ part of the frame
field be reflection symmetric. We are not aware of a complete
Hamiltonian treatment in absence of this (or analogous) condition.

\section{Action and the Covariant Phase space}
\label{s2}

Our basic gravitational variables will be co-frames $e_a^I$ and
Lorentz connections $A_a^{IJ}$ on space-time $\M$. Co-frames $e$
are `square-roots' of metrics and the transition from metrics to
frame fields is motivated by the fact that these frames are
essential if one is to introduce spinorial matter. $e_a^I$ is an
isomorphism between the tangent space $T_p(\M)$ at any point $p$
and a fixed internal vector space $V$ equipped with a metric
$\eta_{IJ}$ with Lorentzian signature $(-+..+)$. The internal
indices can be freely lowered and raised using this fiducial
$\eta_{IJ}$ and its inverse $\eta^{IJ}$. Each co-frame defines a
space-time metric by $g_{ab}:= e_a^I e_b^J \eta_{IJ}$ which also
has signature $(-+..+)$. Then the co-frame $e$ is automatically
orthonormal with respect to $g$. Since the connection 1-forms $A$
take values in the Lorentz Lie algebra, $A_a^{IJ} = -A_a^{JI}$.
The connection acts only on internal indices and defines a
derivative operator
\[
D_a k_I :=\partial_a k_I + {A_{aI}}^J k_J \, ,
\]
where $\partial$ is a fiducial derivative operator which, as usual,
will be chosen to be torsion-free and compatible with $\eta_{IJ}$.
As fundamental fields, $e$ and $A$ are independent. However, the
equation of motion of $A$ implies that $A$ is compatible with $e$,
i.e., is fully determined by $e$. Therefore, boundary conditions on
$A$ will be motivated by those on $e$.

In the Lagrangian and Hamiltonian frameworks we have to first
introduce the precise space of dynamical fields of interest. Let us
fix, once and for all, a co-frame ${}^o\!e_a^I$ such that $g^o_{ab}
= \eta_{IJ}\, {}^o\!e_a^I\, {}^o\!e_b^J$ is flat and
$\partial_{[a}\, {}^o\!e_{b]}^I=0$.  The cartesian coordinates $x^a$
of $g^o_{ab}$ and the associated radial-hyperboloid coordinates
$(\rho, \Phi^i)$ will be used in asymptotic expansions near spatial
infinity.

Detailed analysis shows that to define the Lorentz angular momentum
$e_a^I$ has to admit an expansion to order $d-2$ (see Sec.
\ref{s3.2}). Therefore, we will assume that $e_a^I$ can be expanded
as follows: Setting $n=d-3$,

\be\label{e} e = {}^o\!{e}(\Phi) + \f{{}^{n}e (\Phi)}{\rho^{n}} +
\f{{}^{n+1}e(\Phi)}{\rho^{n+1}} + o(\f{1}{\rho^{n+1}}) \ee
where the leading non-trivial term, ${}^n\!e(\Phi)$, is assumed to
be reflection symmetric and where the remainder $o(\rho^{-m})$ has
the property that $\lim_{\rho \rightarrow \infty}\,\,
\rho^m\,o(\rho^{-m}) =0$.

Since the equation of motion for $A$ implies that $A$ is
compatible with $e$, without any loss of generality we can require
that, in the expansion near spatial infinity, $A_a^{IJ}$ is
completely determined by $e_a^I$ to appropriate leading orders.
This leads us to require that $A_a^{IJ}$ have the following
asymptotic behavior:
\be\label{A} A = {}^{o}\!{A}(\Phi) + \f{{}^{1}\!{A}(\Phi)}{\rho} +
... + \f{{}^{n+2}\!{A}(\Phi)}{\rho^{n+2}} + o(\rho^{n+2}) \ee
where ${}^{0}\!{A} =...= {}^{n}\!{A} = 0$ and ${}^{n+1}\!A$ is given
by
\be \label{A2} {}^{n+1}\!{A_{a}^{IJ}}(\Phi) = 2\rho^{n+1}\,\,
\partial^{[J}\,\left( (\rho^{-n})\,\, {}^{n}\!e_{a}^{I]} \right)
\ee
(In spite of the explicit factors of $\rho$ the right side is in
fact independent of $\rho$ because $\p_a\, {}^n\!e\,\, \sim\,\,
\rho^{-1}\times$ (angular derivatives of\, ${}^n\!e$).) We will
not need the corresponding expression of\, ${}^{n+2}\!A$\, in
terms of $e$ and therefore demand compatibility between $A$ and
$e$ only via (\ref{A2}). Appendix \ref{a1.1} shows that the
boundary conditions are readily satisfied by the $d$-dimensional
Schwarzschild solution.

\subsection{Action Principle}
\label{s2.1}

Consider as before the $d$-manifold $\M$ bounded by space-like
surfaces $M_1$ and $M_2$. In this section we will restrict
ourselves to the case $d > 4$; for the case $d=4$ see \cite{aes}.
We will consider smooth histories $(e,A)$ on $M$ such that $(e,A)$
are asymptotically flat in the sense specified above, and are such
that $M_1,M_2$  are Cauchy surfaces with respect to the space-time
metrics $g$ defined by $e$, and the pull-back of $A$ to $M_1,M_2$
is determined by the pull-back of $e$. The last condition is
motivated by the fact that, since the compatibility between $e$
and $A$ is an equation of motion, boundary values where this
compatibility is violated are not of interest to the variational
principle. Finally it is convenient to partially fix the internal
gauge on the boundaries. We will fix a constant, time-like
internal vector $n^I$ so that $\p_a n^I =0$ and require that the
histories be such that $n^a := n^I e^a_I$ is the unit normal to
$M_1$ and $M_2$.

The first order gravitational action on these histories is given
by (see e.g. \cite{afk,klp,apv})
\begin{equation}\label{action}
S (e, A ) =
    -\,\frac{1}{2\k} \int_{\M} \Sigma^{IJ}\wedge F_{IJ}
   \, +\,\frac{1}{2\k} \int_{\p\M} \Sigma^{IJ} \wedge A_{IJ}
     \, ,
\end{equation}
where the $d$-2-forms $\Sigma^{IJ}$ are constructed from the
co-frames and $F$ is the curvature $A$:
\[
\Sigma_{IJ}:= \textstyle{\f{1}{(d-2)!}}\,\epsilon_{IJK...L} e^K
\wedge ... \wedge e^L \quad\quad {\rm and} \quad\quad F_{I}{}^{J}
= \dd A_{I}{}^{J} + A_{I}{}^{K} \wedge A_{K}{}^{J} \qquad
 \, .
\]
As in more familiar field theories, the action now depends only on
the fundamental fields and their first derivatives. Although the
connection $A$ itself appears in the surface term at infinity,
action is in fact gauge invariant. Indeed, it is not difficult to
show that the compatibility between the pull-backs to $M_1$ and
$M_2$ of $e$ and $A$ and the property $\partial_a n^I=0$ implies
that, on boundaries $M_1$ and $M_2$, $\Sigma^{IJ} \wedge A_{IJ} =
2 K\,\, {}^{d-1}\!\epsilon$ where $K$ is the trace of the
extrinsic curvature of $M_1$ or $M_2$ and ${}^{d-1}\!\epsilon$ is
the volume element thereon (see e.g., section 2.3.1 of
\cite{alrev}). Thus, on $M_1$ and $M_2$, the surface term in
(\ref{action}) is \emph{precisely the Gibbons-Hawking surface
term} with $C=0$ in (\ref{gh}). Therefore, these surface
contributions are clearly gauge invariant. This leaves us with
just the surface term at the time-like cylinder $\tau_{\infty}$ at
infinity. However, since $e$ has to tend to the fixed co-frame
$\ez$ at infinity, permissible gauge transformations must tend to
identity on $\tau_{\infty}$. Since the surface integral on
$\tau_{\infty}$ involves only the pull-back of $A$ to
$\tau_{\infty}$, it follows immediately that this surface integral
is also gauge invariant. Note also that on $\tau_\infty$ this term
is \emph{not} equal to the Gibbons-Hawking surface term (because
$\p_a \p_b\rho$ falls off only as $1/\rho$). Therefore, even if we
were to assume compatibility between $e$ and $A$ everywhere and
pass to a second order action, (\ref{action}) would not reduce to
the Gibbons-Hawking action with $C=0$. It is also inequivalent to
the Gibbons-Hawking prescription of setting $C= K_o$ because, as
we now show, (\ref{action}) is well-defined although it does not
make any reference to an embedding in flat space.

Our boundary conditions allow us to rewrite this action as
\be S(e,A) = \f{1}{2 \kappa} \int_{M} \dd\Sigma \wedge A - \Sigma
\wedge A\wedge A \ee
Boundary conditions also imply that the integrand falls off as
$\rho^{4-2d}$. Since the volume element on any Cauchy slice goes
as $\rho^{d-2}\,\dd\rho\, \dd^{d-2}\Phi$, the action is manifestly
finite \emph{even off shell} if the two Cauchy surfaces $M_1$,
$M_2$ are asymptotically time-translated or even boosted with
respect to each other. Such space-times $\M$ are referred to as
\emph{cylindrical slabs} and \emph{boosted slabs}, respectively.%
\footnote{In specifying our boundary conditions on $e,A$, we set
$n=d-3$. If we restrict ourselves only to the issue of finiteness
of the action, we can weaken these conditions. The action is
finite if $n > \f{d-2}{2}$ in the case of boosted slabs, and $n>
\f{d-3}{2}$ for cylindrical slabs. We introduced stronger
requirements to ensure that the Hamiltonian framework and
conserved charges are well-defined.}

It is easy to check that the functional derivatives of the action
are well defined with respect to both $e$ and $A$ on our class of
histories. Variation with respect to the connection yields $D
\Sigma =0.$ This condition implies that the connection $D$ defined
by $A$ acts on internal indices in the same way as the unique
torsion-free connection $\nabla$ compatible with the co-frame $e$
(i.e., defined by $\nabla_a\, e_b^I =0$). When this equation of
motion is satisfied, the curvature $F$ is related to the Riemann
curvature $R$ of $\nabla$ by
\[{F_{ab}}^{IJ}= {R_{ab}}^{cd}e_c^I e_d^J\, . \]
Varying the action with respect to $e_a^I$ and taking into account
the above relation between curvatures, one obtains Einstein's
equations $G_{ab} = 0$. Inclusion of matter is straightforward
because the standard matter actions contain only first derivatives
of fundamental fields without any surface terms and the standard
fall-off conditions on matter fields imply that the matter action
is finite on cylindrical or boosted slabs even off shell.

\subsection{Covariant Phase Space}
\label{s2.2}

We will now let $\M$ be $\R^d$. The covariant phase space $\Gamma$
will consist of smooth, asymptotically flat \emph{solutions
$(e,A)$ to field equations} on $\M$. (In contrast to section
\ref{s2.1}, the pull-backs of $(e,A)$ are no longer fixed on any
Cauchy surfaces.) Our task is to use the action (\ref{action}) to
define the symplectic structure $\Omega$ on this $\Gamma$.

Following the standard procedure (see, e.g. \cite{abr}), let us
perform second variations of the action to associate with each
phase space point $\gamma \equiv (e,A)$ and tangent vectors
$\delta_1 \equiv (\delta_1 e, \delta_1 A)$ and $\delta_2 \equiv
(\delta_2 e, \delta_2 A)$ at that point, a 3-form $J$ on $\M$,
called the symplectic current:
\begin{equation}\label{symcur}
J(\gamma; \delta_1,\delta_2)=
   -\, \frac{1}{2\k}\,\,[ \delta_1 \Sigma^{IJ}\wedge \delta_2 A_{IJ}
        -\delta_2 \Sigma^{IJ} \wedge \delta_1 A_{IJ} ].
\end{equation}
Using the fact that the fields $(e,A)$ satisfy the field equations
and the tangent vectors $\delta_1, \delta_2$ satisfy the
linearized equations off $(e,A)$, one can directly verify that
$J(\gamma; \delta_1, \delta_2)$ is closed as guaranteed by the
general procedure involving second variations. Let us now consider
a portion ${\M}^\prime$ of $\M$ bounded by two Cauchy surfaces
$M_1, M_2 $. These are allowed to be general Cauchy surfaces so
${\M}^\prime$ may in particular be a cylindrical or a boosted slab
in the sense of section \ref{s3.1}. Consider now a region
${\mathcal{R}^\prime}$ within $\M^\prime$, bounded by finite
portions ${M}^\prime_1, {M}^\prime_2$ of $M_1$ and $M_2$ and a
time-like cylinder $\tau$ joining $\p {M}^\prime_1$ and $\p
{M^\prime_2}$. Since $dJ=0$, integrating it over
${\mathcal{R}^\prime}$ one obtains
\be \int_{{M}^\prime_1} J + \int_{{M}^\prime_2} J + \int_{\tau} J
=0 \ee
The idea is to take the limit as $\tau$ expands to the cylinder
$\tau_\infty$ at infinity. Suppose the first two integrals continue
to exist in this limit \emph{and} the third integral goes to zero.
Then, in the limit the sum of the first two terms would vanish and,
taking into account orientation signs, we would conclude that
$\int_M J$ is a 2-form on $\Gamma$ which is \emph{independent} of
the choice of the Cauchy surface $M$. This would be the desired
pre-symplectic structure. However, the issue of whether the boundary
conditions ensure that the integrals over Cauchy surfaces converge
and the flux across $\tau_\infty$ vanishes is somewhat delicate and
often overlooked in the literature.%
\footnote{Furthermore, even when such issues are discussed, one
often considers only the restricted action $\Omega(\delta,
\delta_V)$ of the pre-symplectic structure $\Omega$, where one of
the tangent vectors, $\delta_V$, is associated with an asymptotic
symmetry $V^a$ on $\M$ because, as we will see, it is this
restricted action that directly enters the discussion of conserved
quantities. Typically the 3-form integrands of $\Omega(\delta_1,
\delta_V)$ on $\M$ have a better asymptotic behavior than those of
generic $\Omega(\delta_1, \delta_2)$. However, unless
$\Omega(\delta_1,\delta_2)$ is well-defined for all $\delta_1,
\delta_2$, one does not have a coherent Hamiltonian framework and
cannot start constructing conserved quantities.}
If either of these properties failed, we would not obtain a
well-defined symplectic structure on $\Gamma$.

Let us first consider the integral over the time-like boundary
$\tau$. As $\tau$ tends to $\tau_\infty$, the integrand $J_{a..bc}
\epsilon^{a..bc}$ tends to
\be \label{flux} \lim_{\tau \rightarrow \tau_\infty}\,\,
\left(\delta_{[1}\f{{}^{n}{\Sigma}_{a...b}}{\rho^n}\right) \,
\left(\delta_{2]} \f{{}^{n+1}\!{A_c}}{\rho^{n+1}}\right)\,
\epsilon^{a...bc}\, =\, \lim_{\tau \rightarrow \tau_\infty}\,\,
\epsilon_{IJK...L}\,\, {}^{o}\!{e_{a}^{K}}\,... (\delta_{[1}
{}^{n}\!{e_{b}^{L}})\, (\delta_{2]} {}^{n+1}\!{A_{c}^{IJ}})\,
\rho^{-1-2n}\,\,\epsilon^{a...bc}\, \ee
where $\epsilon^{a...bc}$ is the metric compatible ($d$-1)-form on
$\tau$. Since the volume element on $\tau$ goes as
$\rho^{d-1}=\rho^{n+2}$, the integral of the symplectic flux over
$\tau_\infty$ is \emph{zero}.

The next question is whether the integral over $\t{M}_1$ (and
$\t{M}_2$) continues to be well-defined in the limit as we
approach $M_1$ (resp. $M_2$). The leading term is again given by
the integral of (\ref{flux}) over $M_1$, the only difference being
that $\epsilon^{ab...c}$ is now the metric compatible ($d$-1)-form
on $M_1$. Since the volume element on $M_1$ goes as $\rho^{d-2}\,
\dd\rho\, \dd^{d-2}\Phi$, a power counting argument shows that the
integrand of this leading term falls off as $\rho^{3-d}$. Thus,
because of our boundary conditions, we are led to a well-defined
pre-symplectic structure, i.e., a closed 2-form, on $\Gamma$
\be \Omega(\delta_{1},\delta_{2})= \f{1}{2 \kappa}\, \int_{M} {\rm
Tr}\, \left[ \delta_{1} \Sigma \wedge \delta_{2} A - \delta_{2}
\Sigma \wedge \delta_{1} A \right]\, ,\ee
where $M$ is any Cauchy surface in $\M$ and trace is taken over the
internal indices. $\Omega$ is not a symplectic structure because it
is degenerate. The vectors in its kernel represent infinitesimal
`gauge transformations'.  The physical phase space is obtained by
quotienting $\Gamma$ by gauge transformations and inherits a true
symplectic structure from $\Omega$. We will not carry out the
quotient however because the calculation of Hamiltonians can be
carried out directly on $(\Gamma, \Omega)$.

\section{Generators of asymptotic Poincare Symmetries}
\label{s3}

Let ${}^o\!e$ and ${}^o\!e^\prime$ be any two flat co-frames in
the phase space $\Gamma$ and denote the corresponding space-time
metrics by $g^o$ and $g^{o\prime}$. Let $V$ be a Killing vector
field of $g^o$. Then, it is easy to check that $g^{o\prime}$
admits a Killing vector $V^\prime$ such that $\lim{\rho
\rightarrow \infty}\, (V^\prime - V) =0$. Killing vectors of any
of these flat metrics will be referred to as asymptotic
symmetries.

Let $V^a$ be an asymptotic symmetry. Then, at any point $(e,A)$ of
$\Gamma$,\, the pair $(\L_V e, \L_V A)$ of fields satisfies the
linearized field equations, whence $\delta_V := (\L_V e, \L_V A)$ is
a vector field on $\Gamma$. (In the definition of the
Lie-derivative, internal indices are treated as scalars; thus $\L_V
e_a^I = V^b\partial_b e_a^I + e_b^I\,
\partial_a V^b$.) The question is whether $\delta_V$ is a phase
space symmetry, i.e., whether it satisfies $\L_{\delta_V} \Omega
=0$.

Consider the 1-form $X_V$ on $\Gamma$ defined by
\be\label{defx} X_V (\delta)= \Omega(\delta, \delta_{V}). \ee
$\L_{\delta_{V}}\, \Omega = 0$ on $\Gamma$ if and only if $X_V$ is
closed, i.e.,
\[\D X_V =0\]
where $\D$ denotes the exterior derivative on (the infinite
dimensional) phase space $\Gamma$. If this is the case then, up to
an additive constant, the Hamiltonian is given by
\[\D H_V = X_V.\]
The constant is determined by requiring that all Hamiltonians
generating asymptotic symmetries at the phase space point $(\ez,
A=0)$ corresponding to Minkowski space-time must vanish. To
calculate the right side of (\ref{defx}), it is useful to note the
Cartan identities
\be\label{liet}
    \L_V A = V \cdot F + D(V\cdot A) \quad\quad{\rm and} \quad\quad
    \L_V \Sigma = V\cdot D\Sigma + D(V\cdot \Sigma) -[(V\cdot A),\Sigma]
\ee
Using these, the field equations satisfied by $(e,A)$ and the
linearized field equations for $\delta$, one obtains the required
expression of $X_V$ (see e.g. \cite{afk,apv}):
\be\label{xt} X_V(\delta) := \Omega(\delta,\delta_V)=
\,\frac{1}{2\k} \oint_{S_\infty} {\rm Tr}\left[(V\cdot A) \delta
\Sigma + \delta A \wedge (V\cdot \Sigma)\right] \, . \ee
Note that the expression involves integrals \textit{only} over the
$d$-2-sphere boundary $S_\infty$ of the Cauchy surface $M$ (i.e.,
the intersection of $M$ with the hyperboloid $\H$ at infinity);
there is no volume term. This is a reflection of the fact that,
because there are no background fields, all diffeomorphisms which
are asymptotically identity represent gauge transformations.

\subsection{Energy-Momentum}
\label{s3.1}
Let us begin by setting $V^a=T^a$, an infinitesimal asymptotic
translation. Since $\delta \Sigma \sim 1/\rho^{d-3},\,\, A \sim
1/\rho^{d-2}$ and since the area element of the $d$-2-sphere grows
as $\rho^{d-2}$, the first term on the right side of (\ref{xt})
vanishes in the limit and we are left with
\be X_T(\delta) := \Omega(\delta,\delta_{T}) = \f{1}{2\kappa}
\oint_{S_{\infty}} {\rm Tr}\, [\delta A \wedge (T \cdot
{}^{o}\Sigma)] \ee
which is manifestly well-defined. Furthermore, since
${}^{o}\Sigma$ and $T$ are constant on the phase space $\Gamma$,
we can pull the variation $\delta$ out of the integral. The
resulting Hamiltonian $H_T$ generating an asymptotic translation
$T^a$ is then given by:
\be H_T = \f{1}{2\kappa} \oint_{S_{\infty}} {\rm Tr}\,[A \wedge (T
\cdot {}^{o}\Sigma)]\ee
Had we selected a translational Killing field $\bar{T}^a$ of
another flat metric $\bar{\eta}_{ab}$ in our phase space $\Gamma$,
we would have obtained the same answer because
$\lim_{\rho\rightarrow \infty} (\bar{T}^a - T^a) = 0$.

Substituting for the leading order term ${}^{n+1}\!A$ in the
asymptotic expansion (\ref{A2}) of $A$, $H_T$ can be expressed in
terms of the leading order co-frame field ${}^n\!e_{ab} =
{}^n\!e_{a}{}^{J}\, {}^o\!e_{bJ}$:
\ba \label{emom} H_T = \lim_{\rho\rightarrow \infty}\,
\f{1}{\kappa}\, \oint_{S_\rho}&& \Big[ (\rho\cdot T)\, \big( \rho
\,n^b\,\p^a\, ({}^n\!e_{ab}) - n \rho^an^b\,\, {}^n\!e_{ab} - \rho
n^a\p_a\,({}^n\!e_{ab})\big)\nonumber\\
&&+ (n\cdot T)\, \big(n\rho^a \rho^b\,\, {}^n\!e_{ab} - n\,\,
{}^n\!e_a{}^a - \rho \rho^b\, \p^a\,({}^n\!e_{ab})\big)\nonumber\\
&&+ nT^an^b\, ({}^n\!e_{ab}) + \rho T^a \rho^b n^c\, \p_c\,
({}^n\!e_{ab})\Big]\, \dd^{d-2}S_o \ea
where $S_\rho$ is a $d$-2 sphere cross-section of the hyperboloid
$\rho = {\rm const}$, $n^a$ is the unit normal to $S_\rho$ within
the hyperboloid and $\dd^{n}S_o$ is the area element of the
\emph{unit} $d$-2-sphere. The terms with an explicit
multiplicative factor of $\rho$ have well-defined limits because
$\p_a {}^n\!e_{ab}$ falls off as $1/\rho$. Thus, energy-momentum
is determined directly by the leading correction ${}^n\!e_{ab}$ to
the Minkowskian co-frame ${}^o\!e_{ab}$ in the asymptotic
expansion. Our boundary conditions required that ${}^n\!e_{ab}$ be
even under reflection. However, this condition is not needed to
arrive at the expression (\ref{emom}). Indeed, an examination of
the integrand shows that only the even part of ${}^n\!e_{ab}$
contributes to this energy-momentum. This expression is
`universal' in the sense that it holds in all higher dimensions.
In Appendix \ref{a1.2} we show that it reduces to the more special
expression derived in \cite{aes} in 4 dimensions using the
Beig-Schmidt form of the metric.

If $T^a$ is a unit time-translation, we can choose a hyperplane
$M$ which is orthogonal to it (with respect to $\eta_{ab}$) and
let $S_\rho$ be the intersection of the hyperboloids $\rho = {\rm
const}$ with $M$. Then, $\rho\cdot T$ vanishes and the remaining
terms can be easily shown to equal the familiar expression of the
ADM energy:
\be H_T = \f{1}{2\kappa} \oint_{S_\infty} \big(\underline{\p}_a
q_{bc} - \underline{\p}_c q_{ab} \big)\, q^{ab}\,\, \dd^{d-2} S^c
\ee
where $q_{ab}$ is the intrinsic physical metric on the Cauchy
surface $M$ and $\underline\p$ is the derivative operator
compatible with a flat metric induced on $M$ by $\eta_{ab}$.

If $T^a$ is a space-translation, we can choose a hyperplane $M$
(w.r.t. $\eta_{ab}$) to which $T^a$ is tangential and let $S_\rho$
be the intersection of the hyperboloids $\rho = {\rm const}$ with
$M$. Now, the pull-back $A_{\underline{a}}{}^{IJ}$ to any Cauchy
surface $M$ of the Lorentz connection $A_a{}^{IJ}$ satisfies
$A_{\underline{a}}{}^{IJ}\, n^J = K_a{}^I$ where $n^a := e_a^J\,
n_J$ is the unit normal to $M$ and $K_{ab} := K_a{}^J e_{bJ}$ is
the extrinsic curvature of $M$. (See, e.g., Section 2.3.1 of
\cite{alrev}.) Therefore, the expression of the Hamiltonian
$H_{\vec{T}}$ generating a spatial translation simplifies to the
more familiar form:
\be H_{\vec{T}} = \f{1}{\kappa}\,\oint_{S_\infty} (K_{ab} - K
q_{ab})\, T^a\,\, \dd^{d-2}S^b \ee

\subsection{Relativistic Angular Momentum}
\label{s3.2}

Let us now set $V^a= L^a$, an infinitesimal asymptotic Lorentz
symmetry. For definiteness, we will assume that it is a Lorentz
Killing field of $\eta_{ab}:= \eta_{IJ}\, \ez^I_a \ez^J_b$ so that
it is tangential to the $\rho = {\rm const}$ hyperboloids $\H$.

The question is whether the vector field $\delta_L$ on $\Gamma$ is
Hamiltonian. Let us begin by examining the 1-form $X_L$ on $\Gamma$.
Using (\ref{xt}), we have:

\be \label {xL} X_{L}(\delta) := \Omega(\delta, \delta_L) =
\f{1}{2 \kappa}\, \lim_{\rho \rightarrow \infty}\,
\oint_{S_{\rho}}\, {\rm Tr}\, [(L \cdot A)\, \delta \Sigma\, +\,
\delta A \wedge (L \cdot \Sigma)  ]\ee
where $S_\rho$ is a $d$-2-sphere, the intersection of the $\rho =
{\rm constant}$ hyperboloid $\H_\rho$ with the Cauchy surface $M$
used to evaluate the symplectic structure. Now, as $\rho$ tends to
infinity, $A \sim \rho^{d-2},\, \delta A \sim \rho^{d-2},\, \Sigma
\rightarrow {}^{o}\Sigma,\, \delta \Sigma \sim \rho^{d-3}$ and $L
\sim \rho$. Therefore for $d>4$, the first term vanishes in the
limit. However, the second term in (\ref{xL}) is potentially
divergent. It is here we use the parity condition on ${}^n\!e$:
Using the fact that this field is even under reflections, one can
show that the potentially divergent term in fact vanishes.
Furthermore, since the integral of $A \wedge \L\cdot \delta\Sigma$
vanishes in the limit, we can take the variation $\delta$ out of
the integral and obtain the Hamiltonian $H_L$ representing the
component of the relativistic angular momentum along $L$:
\be \label{J1} H_{L}=
\f{1}{2\kappa}\,\lim_{\rho\rightarrow\infty}\,\oint_{S_\rho} {\rm
Tr}\,(A\wedge {L} \cdot {\Sigma})\, , \ee
To simplify further, one can use the form (\ref{A2}) of the
leading order piece ${}^{n+1}\!A$ of the connection and using
reflection symmetry of ${}^n\!e$ show that its contribution to
(\ref{J1}) vanishes. Therefore, we have:
\be \label{J2} H_{L}= \f{1}{2\kappa}\, \oint_{S_\infty} {\rm Tr}\,
({}^{n+2}\!{A} \wedge \h{L} \cdot {}^{o}{\Sigma})\, , \ee
where $\h{L}^a = L^a/\rho$ is the Lorentz Killing field on the
\emph{unit} hyperboloid $(\H, h^o_{ab})$. As in 4 dimensions
\cite{aes}, in contrast to the energy momentum the angular
momentum is not determined by the leading order deviation of
$(e,A)$ from the ground state $(\ez, A=0)$, but by sub-leading
terms.

Finally, if $L^a$ is a spatial rotation $\phi^a$, we can recast
(\ref{J1}) in a more familiar form. In this case, only the part
$A_{\underline a}^{IJ}n_J$ of the connection contributes to the
integral, where $n^a = n^J e^a_J$ is the normal to the Cauchy
surface $M$, chosen such that $\phi^a$ is tangential to it, and
the underbar below $a$ denotes that this index is pulled back to
$M$. As in the case of spatial momentum, one notes the relation
between the Lorentz connection $A_a{}^{IJ}$ and the extrinsic
curvature $K_{ab}$ on $M$ to rewrite (\ref{J1}) as:
\be H_\phi = \f{1}{\kappa}\,\lim_{\rho\rightarrow\infty}
\oint_{S_\rho} (K_{ab} - K q_{ab})\, \phi^a\,\, \dd^{d-2}S^b \ee

\section{Discussion}
\label{s4}

In this paper we have shown that in the first order formalism
based on co-frames and Lorentz connections, the Lagrangian and
Hamiltonian frameworks can be constructed without having to
introduce an infinite counter term subtraction in the action. The
analysis was considerably simpler than the four dimensional case
\cite{aes} because of absence of logarithmic translations and
super translations in higher dimensions.

For simplicity, in this paper we focused on vacuum Einstein's
equations. However, inclusion of standard matter ---in particular,
scalar, Maxwell and Yang-Mills fields--- with standard boundary
conditions used in Minkowski space is straightforward. There are
no surface terms in the action associated with matter. Similarly,
the expressions of the Hamiltonians generating asymptotic
Poincar\'e transformations are the same as the ones we found in
section \ref{s3}. In particular, the Hamiltonians consist entirely
of $d$-2-sphere surface integrals at spatial infinity and their
integrands do not receive any explicit contributions from matter.
Matter makes its presence felt through constraint equations which,
in response to matter, modify the asymptotic gravitational fields.

\section*{Acknowledgment:} We would like to thank Jonathan Engle and
Don Marolf for discussions. This work was supported in part by the
NSF grant PHY-0456913, the Alexander von Humboldt Foundation and the
Eberly research funds of Penn State.

\begin{appendix}
\section{Examples}
\label{a1}
\subsection{The Schwarzschild Metric}
\label{a1.1}

In this Appendix we show that $d$-dimensional Schwarzschild
space-times satisfy our boundary conditions.

Recall that the schwarzschild line element can be expressed as
\be \dd s^2 = -(1-\f{2M}{r^{d-3}}) \dd t^2 +
(1-\f{2M}{r^{d-3}})^{-1} \dd r^2 + r^2 \dd\Omega^2 \ee
Where $\Omega$ is the unit solid angle. We can recast this
expression in terms of our variables as:
\be \dd s^2 = {}^o\!\dd s^2 + \f{{}^n\dd s^2}{\rho^n} + O(\rho^-2)
\ee
where ${}^o\dd s^2$ is the flat (Minkowski) metric adapted to the
$(t, r, \Phi^i)$ coordinates and
 \be {}^n\dd s^2 = \f{2M}{\cosh(\chi)}\left[(\sinh^2{\chi} +
\cosh^2{\chi}) \dd\rho^2 + 4\rho \cosh(\chi) \sinh(\chi) \dd\rho
\dd\chi + \rho^2 (\sinh^2{\chi} + \cosh^2{\chi}) \dd\chi^2\right]
\ee
Note that the leading-order deviation from the flat metric
contains is an off-diagonal, $\dd\rho \dd\chi$ term. It is often
assumed (see, e.g. \cite{mm,mmv}) that this off-diagonal term is
absent because we know from general considerations that it is
possible to eliminate it by a suitable redefinition of coordinates
\cite{bs}. However, it is not trivial to carry out this step
\emph{explicitly} even for the Schwarzschild metric. But it is
easy to construct a co-frame compatible with this metric
satisfying the boundary conditions of section \ref{s2}:
\be \label{coframe} {}^n\! e_a^I = \f{M}{\cosh(\chi)}
\left[(\sinh^2{\chi} + \cosh^2{\chi}) \rho_a \rho^I + \f{4}{\rho}
\sinh(\chi) \cosh(\chi) \rho_a \chi^I +\f{1}{\rho^2}
(\sinh^2{\chi} + \cosh^2{\chi}) \chi_a \chi^I \right] \ee
The boundary condition on the leading-order, non-trivial
contribution, ${}^{n+1}\!A_a^{IJ}$ of the connection $A_a^{IJ}$ is
readily satisfied because the connection is \emph{everywhere}
compatible with the co-frame.

Inserting the expression (\ref{coframe}) of the co-frame into our
expression (\ref{emom}) for energy-momentum and (\ref{J2}) of
angular momentum we find that we recover the expected result: $E =
M,\,\, \vec{P}\cdot \vec{T} = 0,\,\, H_L= 0$.

\subsection{4-dimensions: Beig-Schmidt form}
\label{a1.2}

In 4 dimensions, one makes an extensive use of a the `Beig-Schmidt
form' of the metric:
\be  \dd s^{2}=(1+\f{2\sigma} {\rho}) \dd\rho^{2}\, +\,
(1-\f{2\sigma}{\rho})\, \rho^{2}\,\, h_{ab} \dd x^{a} \dd x^{b}
\,+\, o(\rho^{-1}) \ee
where $h_{ab}$ is the metric on the unit hyperboloid. It is easy
to construct a co-frame compatible with this metric by setting:
\be {}^1\! e_a^I = \sigma (2\rho_a \rho^I - {}^o\! e_a^I) \ee
In \cite{aes}, this asymptotic form of the co-frame was used to
obtain an expression for energy-momentum in terms of $\sigma$ and
angular momentum in terms of the sub-leading term ${}^3\!A$ in the
expansion of the Lorentz connection. In this paper, on the other
hand, we have obtained more general forms of these conserved
quantities without assuming the Beig-Schmidt form of the metric.
Do these more general forms directly reduce to those obtained in
\cite{aes} in 4 dimensions once the co-frames are assumed to admit
the Beig-Schmidt form? The answer is in the affirmative: our
expression (\ref{emom}) of energy-momentum and (\ref{J2}) of
angular momentum of section \ref{s3} directly simplify to yield
\be E = \f{2}{\kappa} \oint_{S_\infty} \sigma \dd S_o \ee
\be \vec{P}\cdot \vec{T}= \f{2}{\kappa} \oint_{S_\infty}
\big(\f{\rho \cdot T}{\rho}\big)\,\, \f{\partial\sigma}
{\partial\chi}\, \dd S_o \ee
\be J_{L}= \f{1}{2\kappa}\oint_{S_\infty} (L \cdot
{}^{0}{\Sigma_{IJ}}) \wedge {}^{3}\!A^{IJ}\, . \ee
These are exactly the same expressions that we found in
\cite{aes}. Note, however, that in 4-dimensions the Beig-Schmidt
form was essential to eliminate ambiguities arising from the
logarithmic translations and super translations and to ensure that
the symplectic structure is well-defined. It is just that the
final expressions of conserved quantities have the same
`universal' form that we found in this paper for higher
dimensions.

\end{appendix}

\end{document}